\documentclass[aps,reprint,longbibliography]{revtex4-1}
\usepackage{xcolor}
\usepackage{pdfcolmk}
\usepackage{amsbsy}
\usepackage{amstext}
\usepackage{amssymb}
\usepackage{graphicx}
\usepackage{wasysym}
\usepackage[unicode=true,pdfusetitle,
 bookmarks=true,bookmarksnumbered=true,bookmarksopen=false,
 breaklinks=true,pdfborder={0 0 1},backref=true,colorlinks=false]
 {hyperref}

\makeatletter
\providecommand{\tabularnewline}{\\}

\makeatother

\begin{document}

\title{Salt Effects on the Thermodynamics of \\a Frameshifting RNA Pseudoknot under Tension}

\author{Naoto Hori}
\affiliation{Biophysics Program, Institute for Physical Science and Technology, University of Maryland, College Park, MD 20742, USA}
\altaffiliation[Present address: ]{Department of Chemistry, University of Texas, Austin, TX 78712, USA}

\author{Natalia A. Denesyuk}
\affiliation{Biophysics Program, Institute for Physical Science and Technology, University of Maryland, College Park, MD 20742, USA} 

\author{D. Thirumalai}
\email{dave.thirumalai@gmail.com}
\affiliation{Biophysics Program, Institute for Physical Science and Technology, University of Maryland, College Park, MD 20742, USA}
\altaffiliation[Present address: ]{Department of Chemistry, University of Texas, Austin, TX 78712, USA}


\begin{abstract}
Because of the potential link between $-1$ programmed ribosomal frameshifting  
and response of a pseudoknot (PK) RNA to force, 
a number of single molecule pulling experiments have been performed on PKs to decipher the mechanism
of programmed ribosomal frameshifting. Motivated in part by these experiments,
we performed simulations using a coarse-grained model of RNA to describe the response of a
PK over a range of mechanical forces ($f$s) and monovalent salt concentrations
($C$s). The coarse-grained simulations quantitatively reproduce
the multistep thermal melting observed in experiments, thus validating our model. The free energy
changes obtained in simulations are in excellent agreement with experiments.
By varying $f$ and $C$, we calculated the phase diagram that shows a sequence of structural transitions, 
populating distinct intermediate states.
As $f$ and $C$ are changed, the stem-loop tertiary interactions rupture first, followed by unfolding
of the $3^{\prime}$-end hairpin ($\textrm{I\ensuremath{\rightleftharpoons}F}$).
Finally, the $5^{\prime}$-end hairpin unravels, producing an extended
state ($\textrm{E\ensuremath{\rightleftharpoons}I}$). A theoretical
analysis of the phase boundaries shows that the critical force for
rupture scales as $\left(\log C_{\textrm{m}}\right)^{\alpha}$ with
$\alpha=1\,(0.5)$ for $\textrm{E\ensuremath{\rightleftharpoons}I}$
($\textrm{I\ensuremath{\rightleftharpoons}F}$) transition. This relation
is used to obtain the preferential ion-RNA interaction coefficient,
which can be quantitatively measured in single-molecule experiments, as done previously for DNA hairpins. 
A by-product of our work is the suggestion that the frameshift efficiency is likely determined by
the stability of the $5^{\prime}$ end hairpin that the ribosome first
encounters during translation.
\end{abstract}

\maketitle


\section*{INTRODUCTION}

The multifarious roles RNA molecules play in controlling a myriad
of cellular functions \cite{Cech2014} have made it important to understand
in quantitative terms their folding \cite{Tinoco1999JMB,Woodson2005,Thirumalai2005,Chen2008,Woodson2011}
and how they respond to external stresses \cite{Tinoco2006}. Among
them, one of the simplest structural motifs is the RNA pseudoknot (PK),
which is involved in many biological functions. The simplest type,
referred to as H-type PK, consists of two stems connected by two loops
in which one of the stems forms base pairs with the loop of the other.
The PK motif, found in many RNA molecules such as telomerase \cite{Theimer2006},
mRNA \cite{Brierley1989}, ribosomal RNA \cite{Powers1991}, transfer-messenger
RNA, and viral RNA \cite{TenDam1992}, is functionally important \cite{Staple2005}.
For instance, a PK in the human telomerase RNA is essential
for enzyme (a ribonucleoprotein complex) activity \cite{Gilley1999}.
Many pathogenic mutations in the human telomerase RNA are found in the well-conserved
PK region of the RNA, further underscoring the importance of the PK \cite{Gilley1999,Theimer2006}.
The presence of PKs is also important in ribozyme catalysis and inhibition
of ribosome translation.

Recently, there is heightened interest in the biophysics of PK folding
because it plays a crucial role in affecting the efficiency of $-1$
programmed ribosomal frameshifting ($-1$ PRF) \cite{Green2008,Ritchie2012,Kim2014}.
Usually, proteins are synthesized when the ribosome reads the mRNA
code in three nucleotide steps until the stop codon is reached. In
$-1$ PRF, the open reading frame of mRNA being translated within the
ribosome is programmed to be shifted by one nucleotide, and consequently, 
the mRNA becomes nonfunctional or produces an entirely different
protein \cite{Chen2007,Chen2008,Giedroc2009,White2011,Dinman2012}.
 The downstream PK of the mRNA is a roadblock at
the entrance to the mRNA channel on the ribosome, which impedes the
translocation of the ribosome. The ribosome must unfold the PK, presumed
to occur by exertion of mechanical force, to complete translocation.
Because frameshifting efficiency could depend on how the PK responds to force, a number
of single-molecule pulling experiments have focused on PKs \cite{Chen2009,Ritchie2012,DeMessieres2014}.
Several factors could determine $-1$ PRF, as evidenced by the multitude of proposals
based on many experimental studies \cite{Napthine1999,Kontos2001,Green2008,Ritchie2012,Kim1999,Mouzakis2013,Belew2014,Li2014}.
Nevertheless, given that ribosome translocation exerts mechanical
force on the downstream mRNA, it is physically reasonable to assume
that resistance of PK to force should at least be one factor determining
the frameshifting efficiency \cite{Green2008,Tinoco2002}.

\begin{figure}
\includegraphics{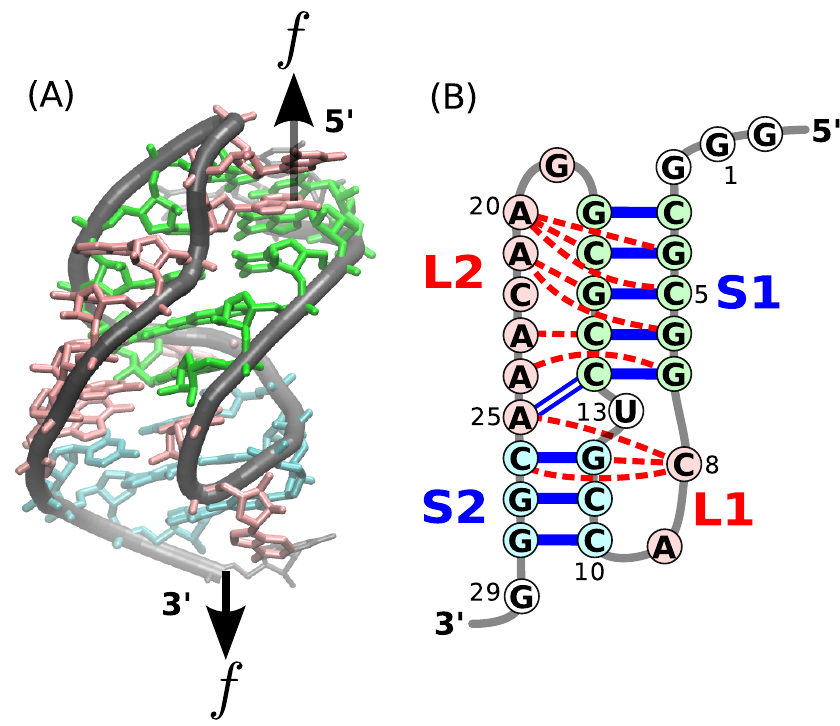}
\caption{\label{fig:Structure}Structure of the BWYV pseudoknot. \textbf{(A)}
Sketch of the crystal structure \cite{Su1999} color-coded according
to secondary structures. The black arrows indicate positions at which
the external mechanical force is applied. \textbf{(B)} Schematic representation
of the secondary structure. Unlike the typical H-type pseudoknot,
the two helix stems (S1 and S2) are not coaxially stacked in the BWYV
pseudoknot \cite{Kim1999}. Tertiary interactions are indicated by red dashed lines.}
\end{figure}

The considerations given above and the availability of both ensemble
and single-molecule pulling experimental data prompted us to focus
on the effects of temperature, mechanical force, and salt concentration
on a 28-nucleotide H-type RNA PK from the beet western yellow virus
(BWYV) \cite{White2011}. This PK has been the subject of several ensemble experiments, which have
characterized the folding thermodynamics as a function of monovalent
and divalent cations \cite{Nixon2000,Soto2007}. The BWYV PK has two
stems (S1 and S2) connected by two loops (L1 and L2) as shown in Fig.~\ref{fig:Structure}. 
The crystal structure reveals that L2 
forms triplex-type tertiary interactions with S1 \cite{Su1999}. In
another context, it has been shown that such interactions and mutations
affecting them dramatically change the efficiency of frameshifting \cite{Chen2009}.
\textit{In vitro} and \textit{in vivo} experiments on many mutants of BWYV suggest variations
in the frameshift efficiency, which, for some mutants, is attributed
to changes in the stem-loop interactions affecting the stability of
PK. For others, direct involvement of the ribosome is invoked \cite{Kim1999}.
In addition to mechanical force ($f$), it is known from several ensemble
experiments that the stability of BWYV PK, like other RNA molecules,
depends on monovalent and divalent cations \cite{Tan2011}. Thus, it is important
to understand in quantitative detail how this crucial structural motif
respond to $f$ and changes in ion concentration.

Although single-molecule pulling experiments suggest how force affects
the stability of RNA \cite{White2011}, structural information obtained
from such experiments is limited only to changes in molecular extensions.
In order to provide a theoretical basis for the changes in the stability
and structural transitions as $f$ and monovalent salt concentration
($C$) are altered, we performed simulations using a coarse-grained
three-site-interaction model (TIS model) \cite{Hyeon2005} of the PK. After
demonstrating that our simulations reproduce the temperature-induced
structural transitions observed in ensemble experiments, we calculated
the global phase diagrams in the $\left[C,f\right]$ plane. The simulations,
performed in the same force range as used in Laser Optical Tweezer
(LOT) experiments \cite{White2011}, quantitatively reproduce the experimental
observations. By combining our simulations and a theory based on ion-RNA
interactions expressed in terms of preferential interaction coefficients \cite{Anderson1993,Soto2007},
we produce several key predictions: (1) The phase diagram in the $\left[C,f\right]$
plane shows that the folded state (F) ruptures in stages populating
distinct intermediates as $f$ increases. Structurally, the low force
intermediate is similar to the one populated during thermal melting.
The sequence of $f$-induced structural transitions can be described
as $\textrm{F\ensuremath{\rightleftharpoons}I\ensuremath{\rightleftharpoons}E}$, 
where $\textrm{E}$ is an extended state; (2) The critical force,
$f_{\textrm{c}}$, to rupture the PK scales as $\left(\log C_{\textrm{m}}\right)^{\alpha}$
with $\alpha=1$ for the $\textrm{I\ensuremath{\rightleftharpoons}E}$
transition and $\alpha=0.5$ for the $\textrm{I\ensuremath{\rightleftharpoons}F}$
transition. We expect this
result to be valid for H-type PKs composed of stems with vastly different
stability, such as the BWYV PK. This result is of general validity
and is applicable to other PKs as well. The slope of the $f_{\textrm{c}}$
versus $\log C_{\textrm{m}}$ is proportional to the changes in the
preferential interaction coefficient
between an intermediate and unfolded states. This prediction can be
used in conjunction with single-molecule experiments allowing for
direct measurement of a key quantity in ion-RNA interactions; and (3)
To the extent the downstream PK is a roadblock for translation, our
work suggests that the ribosome should exert a force $\sim(10-15)\textrm{ pN}$ \cite{Liu2014}
for unimpeded translocation along mRNA, although the actual force value could vary depending on factors 
such as interaction between PK and ribosome. 
Based in part on our work we propose that there should be a link between
the stimulatory effect of frameshifting and the stability of the $5^{\prime}$-end
of the PK. Our predictions are amenable to validations using single-molecule experiments.

\section*{RESULTS AND DISCUSSION}

\subsection*{Temperature-induced transitions in BWYV PK}

In order to validate our model, we first performed temperature replica-exchange
molecular dynamics (REMD) simulations at $C$ = 500 mM with $f$ = 0.
Formations of stems (S1 and S2) and tertiary (stem-loop) interactions
(L1 and L2) are characterized by hydrogen bond (HB) energies $U_{\textrm{HB}}^{\textrm{S}}$
and $U_{\textrm{HB}}^{\textrm{L}}$, respectively. Changes in the
distributions of HB energies as the temperature is increased
are shown in Fig.~\ref{fig:UHB_500mM} in the Supporting Information. Fig.~\ref{fig:UHB_500mM} 
shows that there are two minima corresponding to two distinct states.
The minimum that dominates at $T=20^{\circ}$C corresponds to the
F state in which both S1 and S2 are formed ($U_{\textrm{HB}}^{\textrm{S}}\approx-45\textrm{ kcal/mol}$).
The minimum with $U_{\textrm{HB}}^{\textrm{S}}\approx-30\textrm{ kcal/mol}$
at $\text{80}^{\circ}\textrm{C}$ corresponds to a state in which
S1 is intact (Fig.~\ref{fig:UHB_500mM}). The free energy profile of $U_{\textrm{HB}}^{\textrm{L}}$
shows only one possible minimum around $U_{\textrm{HB}}^{\textrm{L}}\approx-20\textrm{ kcal/mol}$
at $T=20^{\circ}\textrm{C}$, indicating that L1 and L2 fold/unfold
cooperatively as temperature changes.

\begin{figure}
\includegraphics{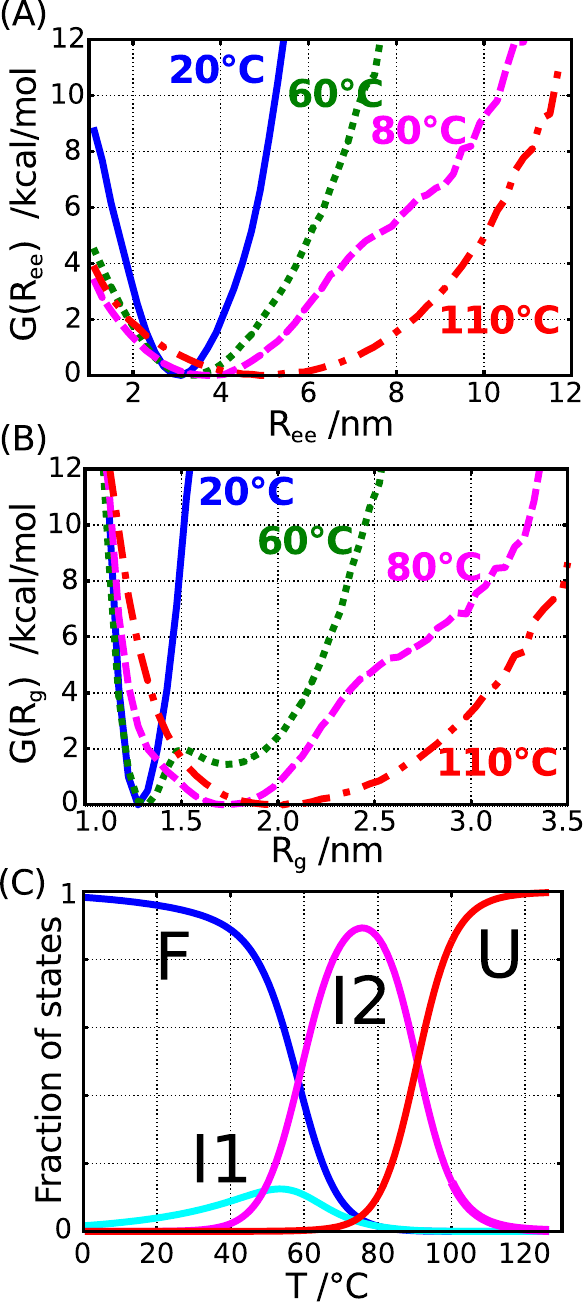}
\caption{\label{fig:TREMD}Characteristics of the thermally induced folding-unfolding
transition of BWYV PK simulated using the temperature-REMD method
at $C=500\textrm{ mM}$ and $f=0$. (A and B) Free energy profiles of
the molecular extension (A; $R_{\textrm{ee}}$) and radius of gyration
(B; $R_{\textrm{g}}$) at 20, 60, 80, and 110$^{\circ}\textrm{C}$.
Each curve is normalized to $G=0$ at its probability $P(R)$ maximum.
(C) Temperature dependence of the fraction of the four prominent states,
which are explicitly labeled.}
\end{figure}

\begin{figure}
\includegraphics{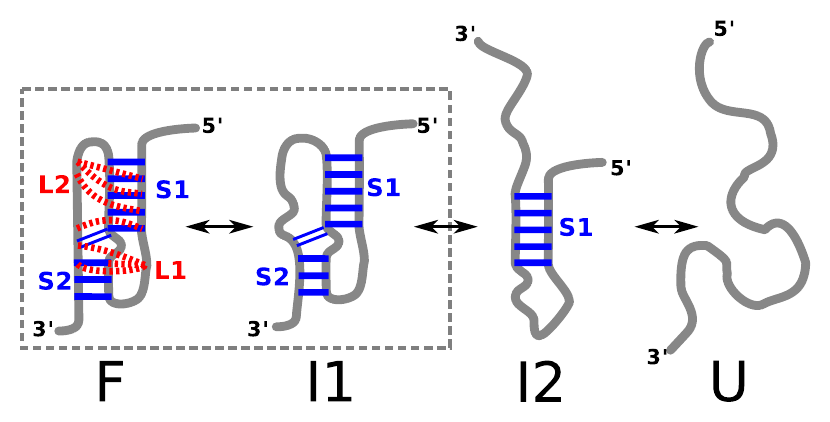}
\caption{\label{fig:schematic}Schematic of the thermodynamic assembly of BWYV
pseudoknot inferred from simulations at zero force. Besides folded
(F) and unfolded (U) states, there are two intermediate state, I1
and I2. The topology of the I1 state is the same as F, except tertiary
interactions are absent.}
\end{figure}

We calculated the free energy profiles $G(R_{\alpha})=-k_{\textrm{B}}T\log P(R_{\alpha})$
($R_{\alpha}=R_{\textrm{\textrm{ee}}}$, end-to-end distance, or $R_{\textrm{g}}$,
radius of gyration) from the distributions of $P(R_{\alpha}),$ which
are given in Fig.~\ref{fig:ProbRgDee_500mM} in the Supporting Information. The locations and the number of
minima report on the thermodynamic transitions between the states
(Fig.~\ref{fig:TREMD}). (1) At a low temperature, $T=20^{\circ}\textrm{C}$,
the PK is completely folded (F state) including all the structural
motifs, S1, S2, L1 and L2. The existence of F state is also confirmed
by the presence of a minimum at $R_{\textrm{ee}}\approx3\textrm{ nm}$
(Fig.~\ref{fig:TREMD}A) and at $R_{\textrm{g}}\approx1.3\textrm{ nm}$
(Fig.~\ref{fig:TREMD}B). These values are similar to those obtained
directly from the crystal structure ($R_{\textrm{ee}}^{\textrm{native}}=2.8\,\textrm{nm}$,
$R_{\textrm{g}}^{\textrm{native}}=1.3\,\textrm{nm}$). (2) At $T=60^{\circ}\textrm{C}$,
the free energy profile for $U_{\textrm{HB}}^{\textrm{L}}$ shows
that some of the tertiary interactions, involving
L1 and L2, are no longer stably formed (Fig.~\ref{fig:UHB_500mM}B). On the other
hand, $G(R_{\textrm{g}})$ shows two stable minima, indicating that
there are two possible states (Fig.~\ref{fig:TREMD}B). In one state,
both stems S1 and S2 are formed but tertiary interactions are absent,
corresponding to the I1 state identified using UV absorbance and differential
scanning calorimetry experiments (Fig.~\ref{fig:schematic}) \cite{Nixon2000,Soto2007}. 
The other conformation, in which only S1
is formed, has been also observed in experiments and is referred to
as the I2 state. It should be noted that the distribution of $U_{\textrm{HB}}^{\textrm{S}}$
also has two minima at $60^{\circ}\textrm{C}$ (Fig.~\ref{fig:UHB_500mM}A) whereas
$G(R_{\textrm{ee}})$ has only one minimum (Fig.~\ref{fig:TREMD}A).
(3) At a still higher temperature $T=80^{\circ}\textrm{C}$, $G(R_{\textrm{ee}})$
and $G(R_{\textrm{g}})$ each have a minimum with S1 stably formed.
This is also reflected in Fig.~\ref{fig:UHB_500mM}, 
which shows a minimum at $U_{\textrm{HB}}^{\textrm{S}}\approx-30\textrm{ kcal/mol}$ 
and a minimum at $U_{\textrm{HB}}^{\textrm{S}}\approx0$. Thus,
completely unfolded conformations, U state, and the I2 state are populated.
(4) At $T=110^{\circ}\textrm{C}$, both $U_{\textrm{HB}}^{\textrm{S}}$
and $U_{\textrm{HB}}^{\textrm{L}}$ are 0 , indicating that only the
U state exists. This is also reflected in $G(R_{\textrm{ee}})$ and
$G(R_{\textrm{g}})$, which show that the PK is completely unfolded.
In both profiles (Figs.~\ref{fig:TREMD}A and~\ref{fig:TREMD}B),
the center of minimum is located at larger values than in the F state
($R_{\textrm{ee}}\approx5\textrm{ nm}$ and $R_{\textrm{g}}\approx2\textrm{ nm}$).

The simulation results in Fig.~\ref{fig:TREMD} and Fig.~\ref{fig:UHB_500mM} show
that melting of BWYV PK undergoes through two intermediate
states, $\textrm{F\ensuremath{\rightleftharpoons}I1\ensuremath{\rightleftharpoons}I2\ensuremath{\rightleftharpoons}U}$
where I1 has the topology of the F state but without significant tertiary interactions (Fig.~\ref{fig:schematic}).
 Although simulations can
be used to distinguish I1 and F states, the intermediate does not
have significant population. The identification of I1 in our simulations
provides a structural support to experiments in which I1 was inferred
from three broad overlapping peaks in the heat capacity curves \cite{Nixon2000,Soto2007}.
It should be stressed that the experimental heat capacity curves exhibit
only two discernible peaks, which would normally imply that there
is only one intermediate present.
The range of melting temperatures for the three transitions can be
obtained from the temperature dependence of the various states $f_{\alpha}(T)$
($\alpha$=F, I1, I2 or U) given in Fig.~\ref{fig:TREMD}C. The
I1 state has a maximum population around $T\approx50^{\circ}\textrm{C}$.
We equate this with the melting temperature $T_{\textrm{m1}}$ associated
with the $\textrm{F\ensuremath{\rightleftharpoons}I1}$ transition.
Similarly, $T_{\textrm{m2}}$ for the $\textrm{I1\ensuremath{\rightleftharpoons}I2}$
transition is $\approx70^{\circ}\textrm{C}$. The $T_{\textrm{m3}}$
for the $\textrm{I2\ensuremath{\rightleftharpoons}U}$ transition
occurs around $90^{\circ}\textrm{C}$ ($f_{\textrm{I2}}(T_{\textrm{m3}})=f_{\textrm{U}}(T_{\textrm{m3}})$).
The values of the three melting temperatures obtained in our simulations
are in remarkable agreement with experiments ($T_{\textrm{m1}}^{\textrm{exp.}}$=$59.4^{\circ}\textrm{C}$,
$T_{\textrm{m2}}^{\textrm{exp.}}$=$69.4^{\circ}\textrm{C}$, and $T_{\textrm{m3}}^{\textrm{exp.}}$=$91.2^{\circ}\textrm{C}$
taken from results of a differential scanning calorimetry experiment at pH=7 in Table 1 of Nixon and Giedroc \cite{Nixon2000}).

The dependence of the stability of the F state with respect to the
U state, $\Delta G_{\textrm{UF}}(T)$ is calculated using a procedure
that does not rely on any order parameter \cite{Denesyuk2013}. The
value of $\Delta G_{\textrm{UF}}(T=37^{\circ}\textrm{C})$ is -14.3
kcal/mol from simulations, which is in excellent agreement with experiments.
The experimental value reported by Nixon and Giedroc for $\Delta G_{\textrm{UF}}(T=37^{\circ}\textrm{C})=-13.3\textrm{ kcal/mol}$, 
whereas Soto et al. estimated that $\Delta G_{\textrm{UF}}(T=37^{\circ}\textrm{C})=-15.1\textrm{ kcal/mol}$ \cite{Nixon2000,Soto2007}.
We also predict that $\Delta G_{\textrm{UF}}(T=25^{\circ}\textrm{C})$
is $-19.0\textrm{ kcal/mol}$. The excellent agreement between simulations
and experiments for the temperature-induced structural transitions
and the $\Delta G_{\textrm{UF}}(T)$ validates the model allowing
us to predict the effect of $f$ on BWYV PK. 

\subsection*{Diagram of states in the $\left[C,f\right]$ plane}

In order to obtain the phase diagram in the $\left[C,f\right]$ plane
of the BWYV PK, we performed a series of low friction Langevin dynamics
simulations by varying the salt concentration from 5 to 1200 mM and the 
mechanical force from 0 to 20 pN, at a constant temperature of $50^{\circ}$C.
We varied $C$ by changing the Debye length in the simulations, and
the mechanical force was externally applied to the terminal nucleotides
(Fig.~\ref{fig:Structure}A).

\begin{figure}
\includegraphics{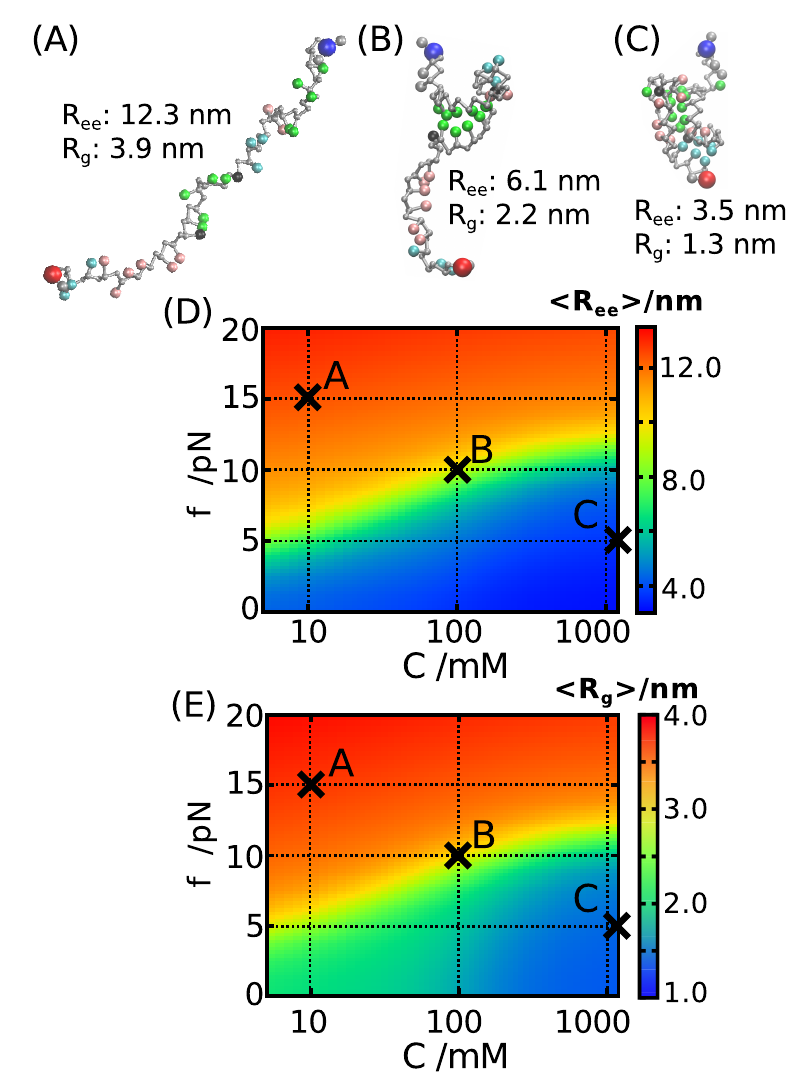}
\caption{\label{fig:Rg_2D}\textbf{(A-C)} Representative snapshots corresponding
to the three distinct states from simulations. \textbf{(D)} Diagram
of states in the $\left[C,f\right]$ plane obtained using extension
$R_{\textrm{ee}}$ as the order parameter. \textbf{(E)} Same as D, 
except $R_{\textrm{g}}$ is used to distinguish the states. The three
crosses in D and E correspond to conditions from which the snapshots
(\textit{A-C}) are sampled. The scale for $R_{\textrm{ee}}$ and $R_{\textrm{g}}$
are given on the right.}
\end{figure}

\begin{figure}
\includegraphics{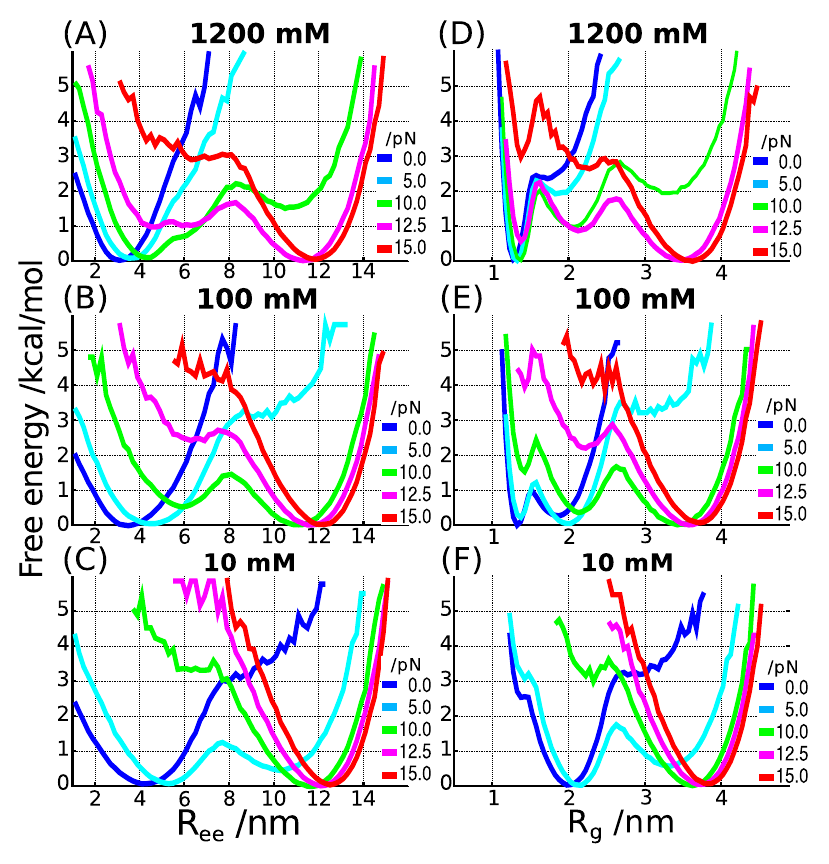}
\caption{\label{fig:Rg_1D}Free energy profiles as a function of $R_{\textrm{ee}}$
\textbf{(A-C)} and $R_{\textrm{g}}$ \textbf{(D-E)} at three different
salt concentrations and a range of forces. The locations of the minima
change as $C$ and $f$ vary, indicating transitions between distinct
states of the BWYV PK.}
\end{figure}

Determination of the phase diagram in the $\left[C,f\right]$ plane
requires an appropriate order parameter that can distinguish between
the various states. In single-molecule pulling experiments, the variable
that can be measured is the molecular extension, $R_{\textrm{ee}}$,
which is conjugated to the applied force. The overall size of the RNA
molecule is assessed using the radius of gyration, $R_{\textrm{g}}$.
Therefore, we characterized the states of the RNA using both $R_{\textrm{g}}$
and $R_{\textrm{ee}}$ as order parameters. The values of the predicted
$R_{\textrm{g}}$ at $f=0$ can be measured using scattering experiments.
Using these two parameters, we obtained the $\left[C,f\right]$ phase
diagram (Fig.~\ref{fig:Rg_2D}). Comparison of the diagram of states
in Figs.~\ref{fig:Rg_2D}D and~\ref{fig:Rg_2D}E reveals common features 
and some differences. From Figs.~\ref{fig:Rg_2D}D and~\ref{fig:Rg_2D}E,
we infer that at $f>12.5\textrm{ pN}$, extended (E) conformations
are dominant at all values of $C$. As the force decreases, the PK
forms compact structures. The boundary separating the extended and
relatively compact phases depends on the salt concentration and 
the value of the mechanical force. The critical force to rupture
the compact structures increases linearly as a function of logarithm
of salt concentration
(Fig.~\ref{fig:Rg_2D}\textit{D}; boundary between red and green
regions). At low forces ($f<2.5\textrm{ pN}$), the diagram of states
based on $R_{\textrm{ee}}$ shows that the extension is relatively
small as $C$ changes from a low to a high value. From this finding, 
one might be tempted to infer that the PK is always folded, which
is not physical especially at low ($C\approx10\textrm{ mM}$) ion
concentrations. In contrast, Fig.~\ref{fig:Rg_2D}E shows that below
5 pN, there is a transition from compact structures ($R_{\textrm{g}}\approx1.3\textrm{ nm}$
in the blue region) at high $C$ to an intermediate state ($R_{\textrm{g}}>2.2\textrm{ nm}$
in the green region) at $C\approx100\textrm{ mM}$. The differences
between the diagrams of states in the $\left[C,f\right]$ plane using
$R_{\textrm{ee}}$ and $R_{\textrm{g}}$ as order parameters are more
vividly illustrated in terms of the free energy profiles $G(R_{\alpha})=-k_{\textrm{B}}T\log(R_{\alpha})$
where $R_{\alpha}$ is $R_{\textrm{ee}}$ or $R_{\textrm{g}}$ (Fig.~\ref{fig:Rg_1D}). 
The profiles $G(R_{\textrm{ee}})$, at three values
of $C$ and different $f$ values, show two minima at most. At $f=0$,
there is only one discernible minimum at $R_{\textrm{ee}}\approx4.2\textrm{ nm}$
at $C=10\textrm{ mM}$. The minimum moves to $R_{\textrm{ee}}\approx3\textrm{ nm}$
at $C=1200\textrm{ mM}$ corresponding to a folded PK. At $f=5\textrm{ pN}$
there are two minima at $C=10\textrm{ mM}$ corresponding to a compact
structure and a stretched conformation (see the cyan profile in Fig.~\ref{fig:Rg_1D}A). 
As $f$ exceeds $5\textrm{ pN}$, there is essentially
one minimum whose position increases as $f$ increases. In the force
regime ($f>5\textrm{ pN}$), only the E state is visible at all $C$.

In order to compare the thermal free energy profiles generated 
at $C$ = 500 mM (Fig. 2), we calculated the $G(R_{\textrm{g}})$ and 
$G(R_{\textrm{ee}})$ at different forces with $C$ fixed at 500 mM. 
Comparison of Fig. 2 and Fig.~\ref{fig:GRgDee_500mM} shows that at $f=0$
the free energy profiles are similar. At $f\neq0$ the folded state
is destabilized, and at $f=15\,\textrm{pN}$, the unfolded state is
more stable than the folded PK.

A much richer diagram of states emerges when the 
$G(R_{\textrm{g}})=-k_{\textrm{B}}T\log P(R_{\textrm{g}})$
is examined. The $G(R_{\textrm{g}})$ profile at $C=10\textrm{ mM}$
is similar to $G(R_{\textrm{ee}})$. However, at higher concentrations, 
there is clear signature of three states (especially at $f=12.5\textrm{ pN}$)
corresponding to the folded PK, an intermediate state, and the E state.
The free energy profiles show that under force, there are minimally
three states, which are F, I, and E. Because $R_{\textrm{ee}}$ cannot
capture the complexity of the states populated as $C$ and $f$ are
varied, it is important to consider $R_{\textrm{g}}$ as well to fully
capture the phase diagram of the PK. We
also computed the free energy profiles as a function of root-mean-square-deviations
(RMSD). In terms of the number and relative positions of basins, the
profiles based on RMSD are qualitatively similar to one of $R_{\textrm{g}}$
(Fig.~\ref{fig:RMSD}).

\subsection*{Formation of stems and loops}

\begin{figure}
\includegraphics{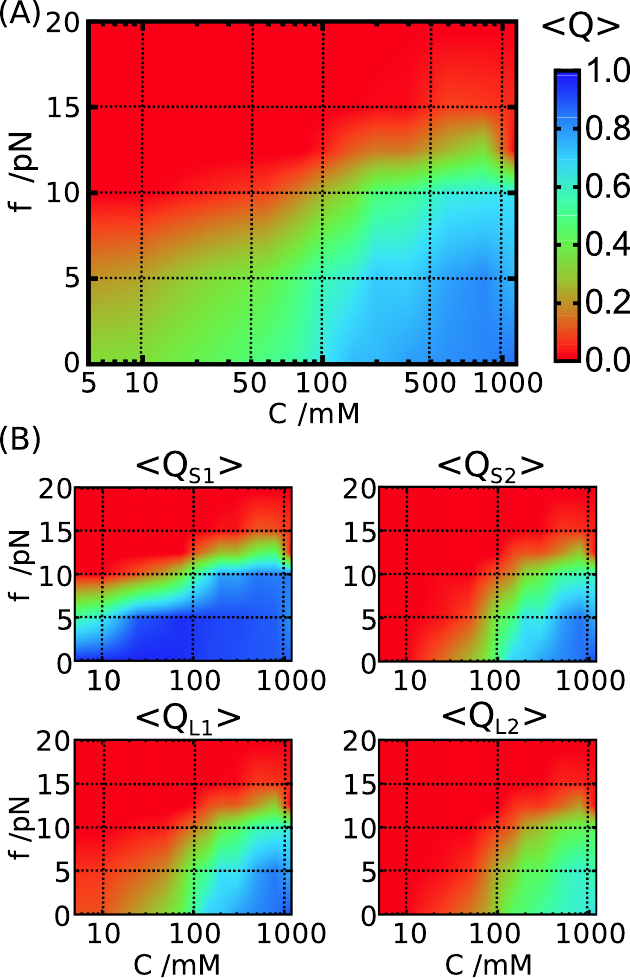}
\caption{\label{fig:Phase-diagram-Q}$\left[C,f\right]$ phase diagram using
the fraction of native contacts as the order parameter. \textbf{(A)}
The diagram of states determined using an average of the total $Q$
for the PK. The scale is given on the right. \textbf{(B)} Decomposition
of $Q$ into stems and loops: S1, stem 1; S2, stem 2; L1, loop 1;
L2, loop 2.}
\end{figure}

In order to highlight the structural changes that occur as $C$ and
$f$ are varied, we used, $Q$, the fraction of native HB
interactions as an order parameter. The $\left[C,f\right]$ phase
diagram (Fig.~\ref{fig:Phase-diagram-Q}A) calculated using the
average $Q$ is similar to the $R_{\textrm{g}}$-phase diagram (Fig.~\ref{fig:Rg_2D}E), 
indicating the presence of three states (Fig.~\ref{fig:Phase-diagram-Q}A ). 
Using $Q$, we can also quantitatively
assess the contributions from different structural elements to the
stability of the PK and correctly infer the order in which the structural
elements of the PK rupture as $f$ is increased. In order to determine
the structural details of each state, we calculated the individual
$Q$s for the two stems ($Q_{\textrm{S1}}$, $Q_{\textrm{S2}}$) and
the two loops ($Q_{\textrm{L1}}$ $Q_{\textrm{L2}}$) (Fig.~\ref{fig:Phase-diagram-Q}B).
The dependence of $\left\langle Q_{\textrm{S1}}\right\rangle $ as
a function of $C$ and $f$ shows that Stem 1 (S1) is extremely stable
and remains intact at all salt concentrations, rupturing only at $f\approx10\textrm{ pN}$
(Fig.~\ref{fig:Phase-diagram-Q}B, upper left panel). In contrast,
the upper right panel in Fig.~\ref{fig:Phase-diagram-Q}B shows
that Stem 2 (S2) is only stably folded above a moderate salt concentration
($C\approx80\textrm{ mM}$) and low $f$. The stability difference
between S1 and S2 can be explained by the number of canonical G-C
base pairs; S1 has five G-C base pairs, whereas only three such pairs
are in S2. Consequently, S1 cannot be destabilized even at low $C$,
but its rupture to produce extended states requires the use of mechanical
force. Above $C>100$ mM, the fraction of contacts associated with
S2 and the two loops ($Q_{\textrm{S2}}$, $Q_{\textrm{L1}}$ and $Q_{\textrm{L2}}$)
simultaneously increase. All the interactions are, however, still
not completely formed ($Q\approx0.8$) even at the highest salt concentration,
$C=1200\textrm{ mM}$. In particular, the contacts involving L2, $Q_{\textrm{L2}}\approx0.6$,
implying that tertiary interactions are only marginally stable at
$T=50^{\circ}\textrm{C}$. 

The results in Fig.~\ref{fig:Phase-diagram-Q} provide a structural
explanation of the transitions that take place as $C$ and $f$ are
changed. (1) In the high force regime, the two stems are fully ruptured
and native interactions are absent creating the E state. (2) Below
$C\approx100$ mM and $f\lesssim10$ pN (left bottom on the $\left[C,f\right]$
plane), only S1 is intact, and thus, a hairpin conformation is dominant.
This is the intermediate state, I2, identified by Soto et al. 
who found that S1 remains stable even at salt concentration as less
as 20 mM NaCl \cite{Soto2007}.
(3) Above $C\gtrsim100$ mM and $f\lesssim10$ pN (right bottom),
both S1 and S2 stems are formed, and the tertiary interactions involving
the two loops are at least partly formed, ensuring that the topology
of the PK is native-like.\textsl{ }At $f=0$ and temperature as a
perturbation, it has been shown that another intermediate state, I1,
exists between I2 and folded states \cite{Nixon2000,Soto2007}. The
I1 state may correspond to conformations in which two stems are fully folded with the entire
topology of the native conformation intact, but tertiary
interactions between stems and loops are incomplete. Although we have
accounted for the presence of such a state in our simulations (see
Fig.~\ref{fig:TREMD}), from here on, we do not distinguish between
I1 and the completely folded state in the phase diagram since the
increases in $Q_{\textrm{S2}}$, $Q_{\textrm{L1}}$ and $Q_{\textrm{L2}}$
on the $\left[C,f\right]$ plane almost overlap. Therefore, we classify
the right bottom region in the $\left[C,f\right]$ phase diagram as
(I1 + F) state. 

\subsection*{Population analysis of the secondary and tertiary interactions}

The $\left[C,f\right]$ phase diagrams provide a thermodynamic basis
for predicting the structural changes that occur as $f$ and $C$
are varied. However, they capture only the average property of the
ensemble within each state without the benefit of providing a molecular
picture of the distinct states, which can only be described using
simulations done at force values close to those used in LOT experiments. 
Therefore, we investigated the distribution
of interactions to ascertain the factors that stabilize the three
states as $C$ and $f$ are varied. Our analysis is based on the HB energy, 
$U_{\textrm{HB}}$. The energy function for HB captures
the formation of base pairs and loop-stem tertiary interactions, taking
into account both distance and angular positions of the donor and
the acceptor \cite{Denesyuk2013}.

\begin{figure}
\includegraphics{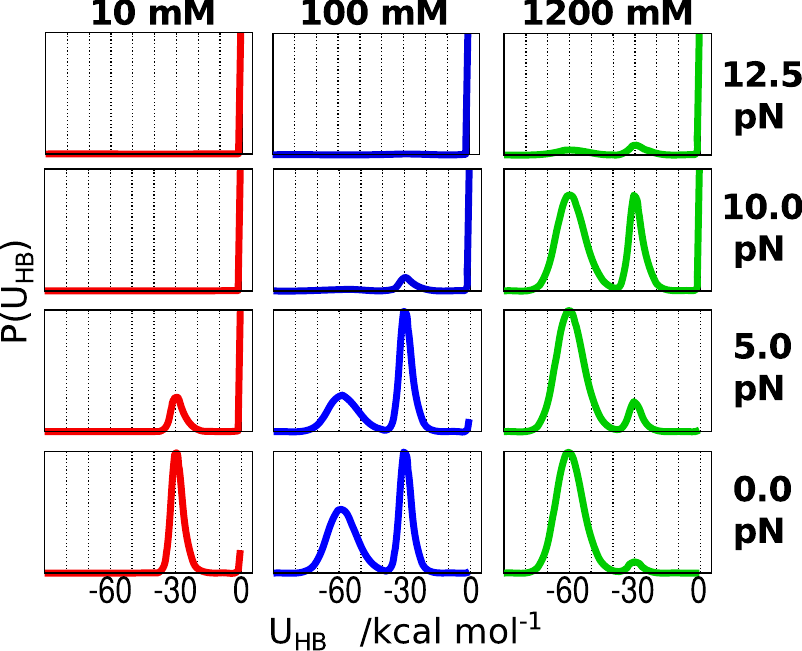}
\caption{\label{fig:HB_2nd}Probability distributions of the total hydrogen
bond energy as a function of $C$ and $f$, which are labeled. The
distinct peaks are fingerprints of the states that are populated at
a specified $C$ and $f$.}
\end{figure}

Fig.~\ref{fig:HB_2nd} shows that the probability densities of the
total HB energy have three peaks at $U_{\textrm{HB}}\thickapprox0$, $-30$, 
and $-60$ kcal/mol. In the high-force regime ($f>12.5\textrm{ pN}$),
there is only one population at all $C$ around $U_{\textrm{HB}}\thickapprox0$,
which obviously corresponds to the fully extended state. The total
HB energy can be decomposed into contributions arising from four distinct
structural motifs in the PK (Fig.~\ref{fig:Prob_HB_2nd}). The peak at $U_{\textrm{HB}}\thickapprox-30$
kcal/mol (Fig.~\ref{fig:HB_2nd}) arises due to the formation of
S1 (compare Figs.~\ref{fig:HB_2nd} and~\ref{fig:Prob_HB_2nd}A). 
The broad peak at $U_{\textrm{HB}}\thickapprox-60$
kcal/mol (Fig.~\ref{fig:HB_2nd}A) is due to the sum of contributions
from S2 and the tertiary interactions. 

Parsing the $U_{\textrm{HB}}$ due to contributions from interactions
among S1, S2, L1, and L2 produces a picture of thermodynamic ordering
as $C$ ($f$) is increased (decreased) from low (high) value to high
(low) value. Low $C$ and high $f$ correspond to the top left region
in the phase diagrams (Figs.~\ref{fig:Rg_2D} and~\ref{fig:Phase-diagram-Q}).
In all conditions, S1 is the first to form as indicated by the presence
of distinct peaks at $C>100\textrm{ mM}$ and $f\lesssim10\textrm{ pN}$
in Fig.~\ref{fig:Prob_HB_2nd}A.
Under more favorable conditions ($f<10\textrm{ pN}$; $C\approx300\textrm{ mM}$
or greater) S2 and tertiary interactions involving L1 are established
(Figs.~\ref{fig:Prob_HB_2nd} B and C). Only at the highest value of $C$ and $f\lesssim5\textrm{ pN}$,
tertiary interactions involving L2 emerge. Interestingly, the order
of formation of the various structural elements en route to the formation
of their folded state follows the stabilities of the individual elements
with the most stable motif forming first. The PK tertiary interactions
are consolidated only after the assembly of the stable structural elements.
This finding follows from our earlier work establishing that the hierarchical
assembly of RNA PKs is determined by the stabilities of individual
structural elements \cite{Cho2009}.

We did not find any condition in which S1 and S2 are completely folded,
but there are no tertiary interactions, $U_{\textrm{HB}}^{\textrm{L}}\approx0$
(i.e., isolated I1 state). Our simulations show that the tertiary interactions
contribute to the loop energy only after the stems are formed. Thus,
the actual population is distributed between I1 and F, and the tertiary
interactions are only marginally stable in the conditions examined
here. Soto et al. suggested that diffuse $\mbox{Mg}^{2+}$ ions play
an important role in stabilizing such tertiary interactions \cite{Soto2007}.
Because $\mbox{Mg}^{2+}$ is not considered here, we cannot determine
the precise interactions stabilizing the I1 state, especially at $f\neq0$.
In the discussion hereafter, for simplicity, we refer to (I1 + F) state as F.

\subsection*{Critical rupture force and preferential ion interaction coefficient}

\begin{figure*}
\includegraphics{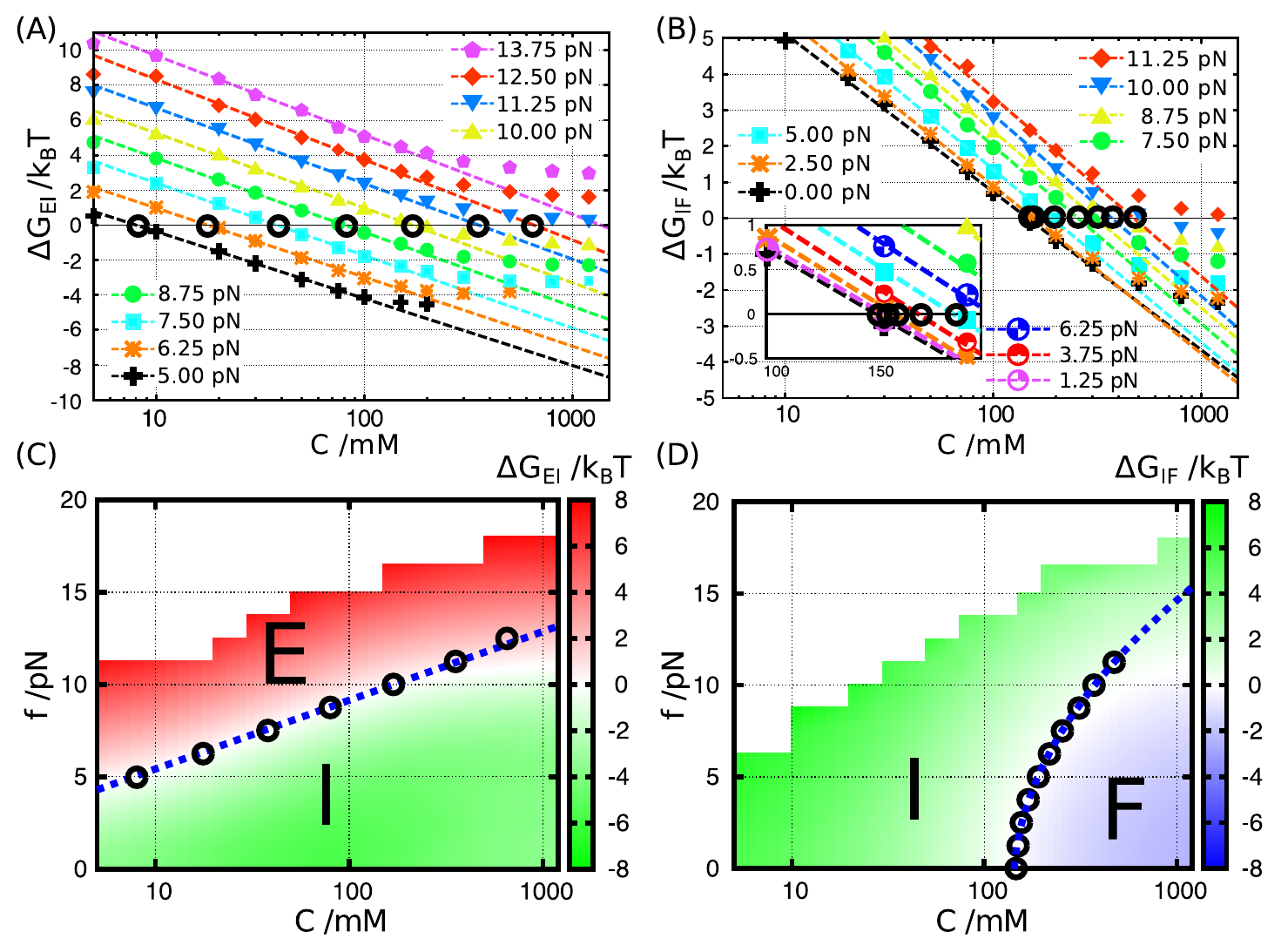}
\caption{\label{fig:dG}Free energy differences between E
and I, and I and F calculated from REMD simulations. The definition of states, E,
I, and F, is based on the hydrogen bond interaction energies associated
with each structural element (see main text for details). The I state
is analogous to I2 in Fig.~\ref{fig:schematic}. \textbf{(A and B)}
Salt concentration dependences of $\Delta G$ is plotted for different
force values (symbols are explained in A and B, respectively). \textbf{(C)}
$\Delta G_{\textrm{EI}}$ is shown on the $\left[C,f\right]$ plane, 
and phase boundary is quantitatively described (white zone between
red and green regions). The blue dashed line is the theoretical prediction
where $f_{\textrm{c}}\propto\log C_{\textrm{m}}$ [Eq.\ref{eq:linear}].
Circles ($\circ$) indicate conditions with $\Delta G_{\textrm{EI}}=0$
extracted from linear dependence of $\Delta G$ on $\log C$ in (A).
It should be noted that the E
state is under tension, and hence is different from the unfolded state
upon thermal melting at $f=0.$ \textbf{(D)} Same as (C) except the
results describe $\textrm{I\ensuremath{\rightleftharpoons}F}$ transition.
Here, the critical force describing the phase boundary is a nonlinear
function of $\log C_{\textrm{m}}$ [Eq.\ref{eq:fc_sqrt}].}
\end{figure*}

In order to delineate the phase boundaries quantitatively, we calculated
the free energy differences between the three major states in the
$\left[C,f\right]$ plane using $\Delta G_{\textrm{EI}}(C,f)=-k_{\textrm{B}}T\log\frac{P(\textrm{I};C,f)}{P(\textrm{E};C,f)}$,
where the classification of the states, E,
I (=I2), or F, is based on the HB interaction energies
of stems, $U_{\textrm{HB}}^{\textrm{S1}}$ and $U_{\textrm{HB}}^{\textrm{S2}}$
(Fig.~\ref{fig:dG}). A similar equation is used to calculate $\Delta G_{\textrm{IF}}$.
Based on the distribution of $U_{\textrm{HB}}$, we determined the
threshold values for S1 formation as $U_{\textrm{HB}}^{\textrm{S1}}<-15\,\textrm{kcal/mol}$
(Fig.~\ref{fig:Prob_HB_2nd}A).
Similarly, S2 formation is deemed to occur if $U_{\textrm{HB}}^{\textrm{S2}}<-10\,\textrm{kcal/mol}$
(Fig.~\ref{fig:Prob_HB_2nd}B).
We classified each structure depending on whether both stems are formed
(F) with both $U_{\textrm{HB}}^{\textrm{S1}}<-15\textrm{ kcal/mol}$
and $U_{\textrm{HB}}^{\textrm{S2}}<-10\textrm{ kcal/mol}$, only S1
is formed (I (=I2)), or fully extended (E).

The classification of the diagram of states based on the calculation
of  $\Delta G_{\textrm{EI}}$
and $\Delta G_{\textrm{IF}}$ is consistent with the $R_{\textrm{g}}$
and $Q$ phase diagrams (Figs.~\ref{fig:dG}C and~\ref{fig:dG}D).
At all values of $f$, $\Delta G$ depends linearly on $\log C$ over
a broad range of $C$ as shown in Figs.~\ref{fig:dG}A and 8B. The
midpoints of salt concentration, $C_{\textrm{m}}$s, at each $f$,
determined using $\Delta G_{\textrm{EI}}(C_{\textrm{m}}^{\textrm{EI}},f)=0$
and $\Delta G_{\textrm{IF}}(C_{\textrm{m}}^{\textrm{IF}},f)=0$, leads
to $f_{\textrm{c}}^{\textrm{EI}}(C_{\textrm{m}}^{\textrm{EI}})$
and $f_{\textrm{c}}^{\textrm{IF}}(C_{\textrm{m}}^{\textrm{IF}})$,
respectively, for the $\textrm{E\ensuremath{\rightleftharpoons}I}$
and $\textrm{I\ensuremath{\rightleftharpoons}F}$ transitions. The
forces $f_{\textrm{c}}^{\textrm{EI}}(C_{\textrm{m}}^{\textrm{EI}})$
and $f_{\textrm{c}}^{\textrm{IF}}(C_{\textrm{m}}^{\textrm{IF}})$
are the critical forces needed to rupture the I and F states, respectively.
The locus of points $f_{\textrm{c}}^{\textrm{EI}}(C_{\textrm{m}})$
and $f_{\textrm{c}}^{\textrm{IF}}(C_{\textrm{m}})$ are shown as circles
in Figs.~\ref{fig:dG}C and~\ref{fig:dG}D, respectively. It is
clear that $f_{\textrm{c}}^{\textrm{EI}}(C_{\textrm{m}}^{\textrm{EI}})$
is linear in $\log C_{\textrm{m}}^{\textrm{EI}}$ along the phase
boundary. In contrast, the dependence of $f_{\textrm{c}}^{\textrm{IF}}(C_{\textrm{m}}^{\textrm{IF}})$
on $\log C_{\textrm{m}}^{\textrm{IF}}$ is nonlinear. In what follows, 
we provide a theoretical interpretation of these results, which automatically
shows that the difference in preferential ion interaction
coefficients between states can be directly measured in LOT experiments \cite{Todd2007,Zhang2008,Dittmore2014,Jacobson2015,Jacobson2016}.

The salt concentration dependence of RNA stability can be determined
using the preferential interaction coefficient, defined as $\Gamma=\left(\frac{\partial C}{\partial C_{\textrm{RNA}}}\right)_{\mu}$, 
where $C_{\textrm{RNA}}$ is the concentration of RNA, and the chemical
potential of the ions, $\mu$, is a constant \cite{Anderson1993,Bond1994}.
The free energy difference, $\Delta G_{\alpha\beta}$ (=$G_{\beta}-G_{\alpha}$),
between two states $\alpha$ and $\beta$ (such as E and I or I and
F) can be expressed as: 
\begin{equation}
\Delta G_{\alpha\beta}\left(C\right)=\Delta G_{\alpha\beta}\left(C_{0}\right)-2k_{\textrm{B}}T\Delta\Gamma_{\alpha\beta}\log\frac{C}{C_{0}},\label{eq:dG(C)}
\end{equation}
where $C_{0}$ is an arbitrary reference value of the salt concentration \cite{Record1998}.
Note that we consider only 1:1 monovalent salt such as KCl or NaCl
for which ion activity can be related to salt concentration $C$.
The factor of 2 in Eq. (\ref{eq:dG(C)}) arises from charge neutrality.
The difference, $\Delta\Gamma_{\alpha\beta}$ (=$\Gamma_{\beta}-\Gamma_{\alpha}$),
is interpreted as an effective number of bound or excluded
ions when the RNA molecule changes its conformation between the
two states \cite{Anderson1993,Bond1994}. The free energy change upon
application of an external force $f$ is $\Delta G_{\alpha\beta}(f)=\Delta G_{\alpha\beta}(0)+\Delta R_{\textrm{ee}}^{\alpha\beta}\cdot f$
where $\Delta R_{\textrm{ee}}^{\alpha\beta}=R_{\textrm{ee}}^{\alpha}-R_{\textrm{ee}}^{\beta}$ \cite{Tinoco2002}.
Thus, 
\begin{eqnarray}
\Delta G_{\alpha\beta}(C,f) & = & \Delta G_{\alpha\beta}(C_{0},0)\nonumber \\
 &  & -2k_{\textrm{B}}T\Delta\Gamma_{\alpha\beta}\log\frac{C}{C_{0}}+\Delta R_{\textrm{ee}}^{\alpha\beta}\cdot f.\label{eq:dG(C,f)}
\end{eqnarray}

In the $\left[C,f\right]$ phase diagram, $\Delta G_{\alpha\beta}$ is 0 
along the phase boundaries. Since the reference salt concentration
$C_{0}$ is arbitrary, we determined its value using $\Delta G_{\alpha\beta}(C_{0},f=0)=0$.
Thus, $C_{0}=C_{\textrm{m},f=0}^{\alpha\beta}$ is the salt concentration
at the midpoint of the transition at zero force. The
determination of $\Delta\Gamma_{\alpha\beta}$ using this procedure
from single-molecule pulling data has been described in several key
papers \cite{Todd2007,Zhang2008,Dittmore2014,Jacobson2015}.
By adopting the procedure outlined in these studies and with our choice of $C_{0}$, we rearrange
Eq.(\ref{eq:dG(C,f)}) to obtain the phase boundary at $f\neq0$, 
\begin{equation}
0=-2k_{\textrm{B}}T\Delta\Gamma_{\alpha\beta}\log\frac{C_{\textrm{m}}^{\alpha\beta}}{C_{\textrm{m},f=0}^{\alpha\beta}}+\Delta R_{\textrm{ee}}^{\alpha\beta}\cdot f_{\textrm{c}}^{\alpha\beta},\label{eq:boundary}
\end{equation}
 where $C_{\textrm{m}}^{\alpha\beta}$ is the midpoint of salt concentration,
and $f_{\textrm{c}}^{\alpha\beta}$ is the critical force associated
with the transition between states $\alpha$ and $\beta$. Measurement of 
 $f_{\textrm{c}}^{\alpha\beta}$ using data from single-molecule pulling experiments has been used to obtain $\Delta\Gamma_{\alpha\beta}$
for few systems that exhibit two-state transitions \cite{Todd2007,Zhang2008,Dittmore2014,Jacobson2015}. Here, we have adopted it for a PK exhibiting more complex
unfolding transitions. The connection between the relation in Eq.
(\ref{eq:boundary}), derived elsewhere, to Clausius-Clapeyron equation
was established recently by Saleh and coworkers \cite{Dittmore2014}.

\begin{figure}
\includegraphics{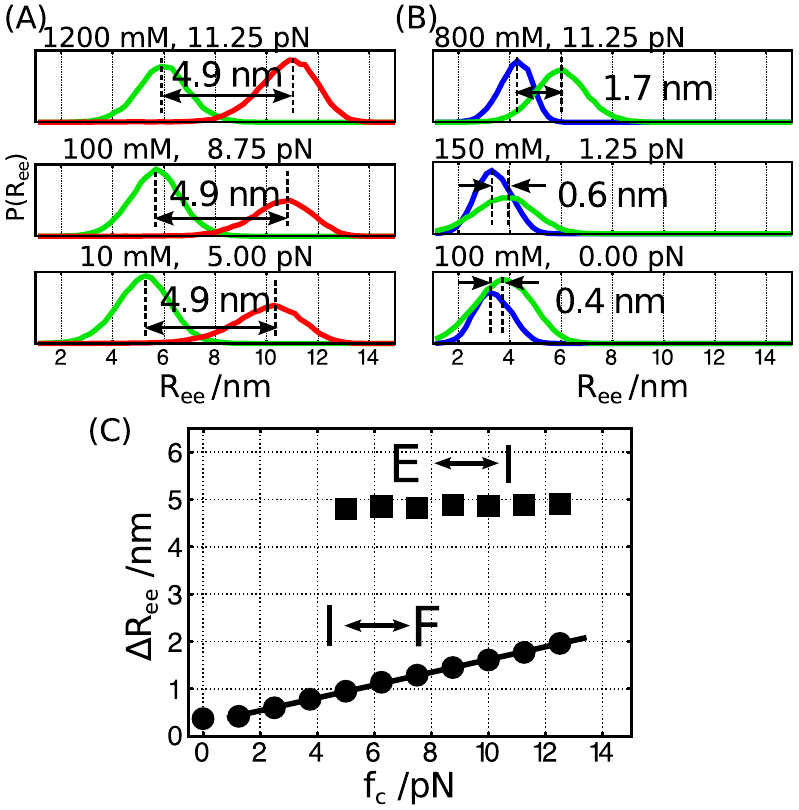}
\caption{\label{fig:dee_states}\textbf{(A and B)} Probability distributions
of the extensions for states E and I (A), and I and F (B). 
All the conditions shown here are near
the phase boundaries in the $\left[C,f\right]$ plane ($\Delta G_{\textrm{EI}}$
and $\Delta G_{\textrm{IF}}$ between the two states are close to
0). Distances between peaks of the two distributions are shown. In
the $\textrm{E\ensuremath{\rightleftharpoons}I}$ transition (A),
the mean distances are almost constant around 4.9 nm. On the other
hand, in the $\textrm{I\ensuremath{\rightleftharpoons}F}$ transition
(B), the mean distance becomes smaller as the mechanical force weakens.
\textbf{(C)} Dependence of $\Delta R_{\textrm{ee}}$ on $f_{\textrm{c}}$
along the phase boundaries. In the $\textrm{I\ensuremath{\rightleftharpoons}F}$
transition, the dependence can be fit using $\Delta R_{\textrm{ee}}^{\textrm{IF}}=af_{\textrm{c}}^{\textrm{IF}}+b$, 
where $a=0.14\textrm{ nm/pN}$ and $b=0.27\textrm{ nm}$ (solid line).}
\end{figure}

It follows from Eq.(\ref{eq:boundary}) that the critical force, 
\begin{equation}
f_{\textrm{c}}^{\alpha\beta}=\frac{2k_{\textrm{B}}T\Delta\Gamma_{\alpha\beta}}{\Delta R_{\textrm{ee}}^{\alpha\beta}}\log\frac{C_{\textrm{m}}^{\alpha\beta}}{C_{\textrm{m},f=0}^{\alpha\beta}},\label{eq:linear}
\end{equation}
 leading to the prediction that there should be a linear dependence
between $f_{\textrm{c}}^{\alpha\beta}$ and the logarithm of the midpoint
of the salt concentration if $\Delta R_{\textrm{\textrm{ee}}}^{\textrm{\ensuremath{\alpha\beta}}}$
is a constant value along the phase boundary. In accord with the prediction
in Eq.(\ref{eq:linear}), we find that $f_{\textrm{c}}^{\textrm{EI}}$
varies linearly with $\log\frac{C_{\textrm{m}}^{\textrm{EI}}}{C_{\textrm{m},f=0}^{\textrm{EI}}}$
along the phase boundary separating E
and I (Fig.~\ref{fig:dG}C) where $\left\langle \Delta R_{\textrm{ee}}^{\textrm{UI}}\right\rangle \approx4.9\,\textrm{nm}$
over a broad range of $f_{\textrm{c}}$ (Figs.~\ref{fig:dee_states}A
and~\ref{fig:dee_states}C). 

The observed nonlinearity of the phase boundary separating I and
F over the entire concentration range (Fig.~\ref{fig:dG}D) arises
because $\Delta R_{\textrm{ee}}^{\textrm{IF}}$ is not a constant 
but varies linearly with $f_{\textrm{c}}^{\textrm{IF}}$ (Figs.~\ref{fig:dee_states}B and~\ref{fig:dee_states}C). 
In giving from $\textrm{F\ensuremath{\leftrightarrow}}\textrm{I}$, there
is only a rupture of few tertiary interactions and unfolding
of the relatively unstable S2. Because this transition is not cooperative
(interactions break in a sequential manner), we believe that a linear
behavior over a limited force range is not unreasonable. In contrast,
the unfolding of S1 is cooperative, like some hairpins studied using
pulling experiments, in the $\textrm{E}\leftrightarrow\textrm{I}$
transition, resulting in the constant value for $\Delta R_{\textrm{ee}}$
over the force range probed. Thus, using $\Delta R_{\textrm{ee}}^{\alpha\beta}=af_{\textrm{c}}^{\alpha\beta}+b$
and Eq.(\ref{eq:linear}), we find that $f_{\textrm{c}}^{\alpha\beta}$ satisfies, 
\begin{equation}
f_{\textrm{c}}^{\alpha\beta}=\frac{2k_{\textrm{B}}T\Delta\Gamma_{\alpha\beta}}{af_{\textrm{c}}^{\alpha\beta}+b}\log\frac{C_{\textrm{m}}^{\alpha\beta}}{C_{\textrm{m},f=0}^{\alpha\beta}}.\label{eq:fc_quadratic}
\end{equation}
Note that if $a$ is zero, then Eq.(\ref{eq:fc_quadratic}) reduces
to Eq.(\ref{eq:linear}) with $b=\Delta R_{\textrm{ee}}^{\alpha\beta}=const.$
Solving Eq.(\ref{eq:fc_quadratic}), the nonlinear dependence of $f_{\textrm{c}}^{\textrm{IF}}$
is expressed as,

\begin{equation}
f_{\textrm{c}}^{\textrm{IF}}=\sqrt{\frac{2k_{\textrm{B}}T\Delta\Gamma_{\textrm{IF}}}{a}\log\frac{C_{\textrm{m}}^{\textrm{IF}}}{C_{\textrm{m},f=0}^{\textrm{IF}}}+\frac{b^{2}}{4a^{2}}}-\frac{b}{2a}.\label{eq:fc_sqrt}
\end{equation}
The simulation data can be quantitatively fit using Eq.(\ref{eq:fc_sqrt})
(Fig.~\ref{fig:dG}D). In general, we expect $\Delta R_{\textrm{ee}}^{\alpha\beta}$
depends on $f_{\textrm{c}}^{\alpha\beta}$, and hence, $f_{\textrm{c}}^{\alpha\beta}\sim\sqrt{\log C_{\textrm{m}}^{\alpha\beta}}$.
It is worth noting that the estimation of $\Delta\Gamma_{\alpha\beta}$
requires only the knowledge of how critical force varies with salt
concentrations ($\Delta G=0$). For any given salt concentration, the
best statistics is obtained at the critical force because at $f_{\textrm{c}}$, 
multiple transitions between the two states can be observed.

From the coefficients in the dependence of $f_{\textrm{c}}$ on $\log C_{\textrm{m}}$,
we can estimate $\Delta\Gamma$ which provides a quantitative estimation
of the extent of ion-RNA interactions \cite{Record1998}. For the $\textrm{E\ensuremath{\rightleftharpoons}I}$
transition, $\left\langle \Delta R_{\textrm{ee}}\right\rangle \approx4.9\,\textrm{nm}$
and the slope of the linear function is 1.7 pN, which leads to $\Delta\Gamma_{\textrm{EI}}=0.96$
(Fig.~\ref{fig:dG}C). This indicates that upon the conformational
change from unfolded to the intermediate state, one ion pair enters
the atmosphere of the PK. For the transition between I and F, fitting the dependence
of $\Delta R_{\textrm{ee}}^{\textrm{IF}}$ on $f_{\textrm{c}}^{\textrm{IF}}$
with $a=0.14\textrm{ nm/pN}$ and $b=0.27\textrm{ nm}$ (Fig.~\ref{fig:dee_states}C),
we estimate $\Delta\Gamma_{\textrm{IF}}=1.8$ using Eq.(\ref{eq:fc_sqrt})
(Fig.~\ref{fig:dG}D). Consequently, the total number of excess
cations effectively bound to folded RNA is expected to be around 2.8,
a value that is in rough accord with the experimental estimate of 2.2 \cite{Nixon2000}.
More importantly, $\Delta\Gamma_{\alpha\beta}$ can be measured using
LOT and magnetic tweezer
experiments \cite{Todd2007,Zhang2008,Dittmore2014,Jacobson2015,Jacobson2016},
thus expanding the scope of such experiments to measure ion-RNA interactions. 
A recent illustration of the efficacy of single-molecule experiments
is the study by Jacobson and Saleh who quantified $\Delta\Gamma$
from force measurements for an RNA hairpin \cite{Jacobson2016}.

Our analysis is based on
the assumption that $\Delta\Gamma$ is a constant independent of the
salt concentration. However, it is known experimentally that $\Delta\Gamma$
deviates from a constant value at high salt concentration \cite{Record1978,Dittmore2014,Jacobson2016}.
In Fig.~\ref{fig:dG}C and D, theoretical lines (blue dotted) show
such a deviation from $\Delta G=0$ of the simulation data (white
region) at high salt concentration ($C\apprge500\,\textrm{mM}$).
Therefore, we believe that our theoretical analysis is most accurate
for $C\lesssim500\,\text{\textrm{mM}}$. Our coarse-grained simulation
reflects the nonlinearity of $\Delta\Gamma$ observed in previous
experiments. Understanding the origin of the nonlinearity requires
a microscopic theory that should take into account the effects of ion-ion correlations.
For complex architectures such as PK or ribozymes, this does not
appear straightforward.

\section*{MODEL AND METHODS}

\subsection*{The three-interaction-site (TIS) model}

We employed a variant of the three-interaction-site (TIS) model, a
coarse-grained model first introduced by Hyeon and Thirumalai for
simulating nucleic acids \cite{Hyeon2005}. The TIS model has been
previously used to make several quantitative predictions for RNA molecules, 
ranging from hairpins to ribozymes with particular focus on folding
and response to $f$ \cite{Hyeon2005,Cho2009,Biyun2011,Lin2013,Denesyuk2015}.
More recently, we have incorporated the consequences of counter ion
condensation into the TIS model, allowing us to predict thermodynamic
properties of RNA hairpins and PK that are in remarkable agreement
with experiments \cite{Denesyuk2013}. In the TIS model, each nucleotide
is represented by three coarse-grained spherical beads corresponding
to phosphate (P), ribose sugar (S), and a base (B). Briefly, the effective
potential energy (for details, see Ref. \cite{Denesyuk2013})
of a given RNA conformation is $U_{\textrm{TIS}}=U_{\textrm{L}}+U_{\textrm{EV}}+U_{\textrm{ST}}+U_{\textrm{HB}}+U_{\textrm{EL}}$,
where $U_{\textrm{L}}$ accounts for chain connectivity and angular
rotation of the polynucleic acids, $U_{\textrm{EV}}$ accounts for
excluded volume interactions of each chemical group, and $U_{\textrm{ST}}$
and $U_{\textrm{HB}}$ are the base-stacking and HB interactions,
respectively. Finally, the term $U_{\textrm{EL}}$ corresponds to
electrostatic interactions between phosphate groups.

The determination of the
parameters for use in simulations of the coarse-grained model of RNA
was described in detail in a previous publication \cite{Denesyuk2013}.
Briefly, we determined the stacking interactions by a learning procedure,
which amounts to reproducing the measured melting temperature of dimers.
We showed that a single choice of model parameters (stacking interactions,
the Debye-H\"{u}ckel
potential for electrostatic interactions, and structure-specific choice
for hydrogen bonds) is sufficient to obtain detailed agreements with
available experimental data for three different RNA molecules. Thus,
the parameters are transferable and we use them verbatim in the present
simulations. The determination of hydrogen bond interaction is predicated
on the structure. Most of these are taken from the A-form RNA structure
and hence can be readily used for any RNA molecule. Only the interaction
lengths and angles in hydrogen-bonding interaction of noncanonical
base pairs are based on the specific PDB structure, which in our case
is the BWYV pseudoknot in this study. We list all the BYWV-PK-specific
values used in our simulations in Table S1.

The repulsive electrostatic interactions between the phosphate groups
is taken into account through the Debye-H\"{u}ckel theory, $U_{\textrm{EL}}=\sum_{i,j}\frac{q^{\ast2}e^{2}}{4\pi\varepsilon_{0}\varepsilon(T)r_{ij}}\exp\left(-\frac{r_{ij}}{\lambda_{\textrm{D}}}\right)$,
where the Debye length is $\lambda_{\textrm{D}}=\sqrt{\frac{\varepsilon(T)k_{\textrm{B}}T}{4\pi e^{2}I}}$.
In the simulations, salt concentration can be varied by changing the
ionic strength $I=\Sigma q_{n}^{2}\rho_{n}$, where $q_{n}$ is the
charge of ion of type $n$, and $\rho_{n}$ is its number density.
Following our earlier study \cite{Denesyuk2013}, we use an experimentally
fitted function for the temperature-dependent dielectric constant
$\varepsilon(T)$ \cite{Malmberg1956}. The renormalized charge on
the phosphate group is $-q^{\ast}e\,(q^{\ast}<1)$. Because of the
highly charged nature of the polyanion, counterions condense onto
the RNA, thus effectively reducing the effective charge per phosphate.
The extent of charge reduction can be calculated using the Oosawa-Manning
theory. Charge renormalization, first used in the context of RNA folding
by Heilman-Miller et al. \cite{Heilman2001thermodynamics} and more
recently incorporated into CG models for RNA by us \cite{Denesyuk2013}
and others \cite{Hayes2014BJ,Hayes2015PRL}, is needed to predict accurately
the thermodynamics and kinetics of RNA folding \cite{Heilman2001kinetics}.
The renormalized value of the charge on the P group is approximately
given by $-q^{\ast}(T)e=\frac{-be}{l_{\textrm{B}}(T)}$, where the
Bjerrum length is $l_{\textrm{B}}(T)=\frac{e^{2}}{\varepsilon(T)k_{\textrm{B}}T}$
and $b$ is the mean distance between the charges on the phosphate
groups \cite{Manning1969}. The mean distance ($b$) between charges
on RNA is difficult to estimate (except for rod-like polyelectrolytes)
because of its shape fluctuations. Therefore, it should be treated
as an adjustable parameter. In our previous study, we showed that $b=4.4\textrm{ \AA}$
provides an excellent fit of the thermodynamics of RNA hairpins and
the MMTV pseudoknot \cite{Denesyuk2013}. We use the same value in
the present study of BWYV as well. Thus, the force field used in this
study is the same as in our previous study attesting to its transferability
for simulating small RNA molecules.

The native structure of the BWYV PK is taken from the PDB (437D) \cite{Su1999}.
The structure contains 28 nucleotides. We added an additional guanosine
monophosphate to both the $5^{\prime}$ and $3^{\prime}$ terminus.
Thus, the simulated PK has 30 nucleotides. The mechanical force is
applied to the ribose sugars of both the $5^{\prime}$- and $3^{\prime}$-
terminus (Fig.~\ref{fig:Structure}A).

\subsection*{Simulation details}

We performed Langevin dynamics simulations by solving the equation
of motion, $m\ddot{\boldsymbol{x}}=-\frac{\partial U_{\textrm{TIS}}}{\partial\boldsymbol{x}}-\gamma\dot{\boldsymbol{x}}+\boldsymbol{R}$,
where $m$ is mass; $\gamma$, the friction coefficient, depends on
the size of the particle type (P, S, or B); and $\boldsymbol{R}$
is a Gaussian random force, which satisfies the fluctuation-dissipation
relation. The $\left[C,f\right]$ phase diagram is calculated at a
somewhat elevated temperature, $T$=$50^{\circ}\textrm{C}$, in order
to observe multiple folding-unfolding transitions when the monovalent
salt concentration is varied over a wide range, from 5 to 1200 mM.
The numerical integration is performed using the leap-frog algorithm
with time step length $\delta t_{L}=0.05\tau$, where $\tau$ is the
unit of time. We performed low friction Langevin dynamics simulations
to enhance the rate of conformational sampling \cite{Honeycutt1992}.
For each condition ($C$ and 
$f$), we performed 50 independent simulations for $6\times10^{8}$
time steps and the last two-third of trajectories are used for data
analyses. In addition, we also performed a set of simulations using
the replica-exchange molecular dynamics (REMD) to confirm the validity
of the results obtained by the low friction Langevin dynamics simulations \cite{Sugita1999}.
In the REMD simulations, both $C$ and $f$ are treated as replica
variables and are exchanged between 256 replicas, by covering $C$
from 1 to 1200 mM and $f$ from 0 to 20 pN. The replicas are distributed
over $16\times16$ grid. The combined set of simulations using different
protocols assures us that the simulations are converged at all conditions.

Our simulations are done
at constant forces under equilibrium conditions in which we observe
multiple transitions between the various states. We have made comparison
to LOT experiments, which are performed by pulling the RNA at a constant
velocity. In general, if RNA is stretched at a constant velocity, then
one has to be concerned about non-equilibrium effects. However, in
optical tweezer experiments, the pulling speed is low enough so that
the tension propagates uniformly throughout the RNA prior to initiation
of unfolding \cite{Hyeon2006}. Thus, the unfolding of PK studied using
LOT experiments \cite{White2011} occurs at equilibrium, justifying
comparisons to our explicit equilibrium simulations.

\subsection*{Analysis of simulation data}

To analyze the structural transitions from which the phase diagrams
are determined, we stored snapshots every $10^{4}$ time steps from
all the trajectories. We computed the radius of gyration ($R_{\textrm{g}}$) 
and the end-to-end distance ($R_{\textrm{ee}}$) using the stored
snapshots. The fraction of native contacts, $Q$, is calculated by
counting the number of hydrogen bond (HB) interaction pairs. The assessment
of HB formation is based on the instantaneous value of the HB interaction
energy, $U_{\textrm{HB}}$. In our model, each HB interaction pair
contributes up to $nU_{\textrm{HB}}^{0}$ toward stability, where
$U_{\textrm{HB}}^{0}$ is -2.43 kcal/mol, corresponding to the stability
of one HB, and $n$ represents number of HB associated
in the interaction \cite{Denesyuk2013}. A cutoff value $nU_{\textrm{HB}}^{\textrm{C}}$
is defined to pick out contacts that are established. We use a value,
$U_{\textrm{HB}}^{\textrm{C}}=-1.15$ kcal/mol to obtain the diagram
using $Q$ as an order parameter. Modest changes in $U_{\textrm{HB}}^{\textrm{C}}$
do not affect the results qualitatively.

\section*{CONCLUDING REMARKS}

Motivated by the relevance of force-induced transitions in PK to PRF, 
we have conducted extensive simulations of
BWYV PK using a model of RNA, which predicts with near quantitative
accuracy the thermodynamics at $f=0$. The phase diagram in the $\left[C,f\right]$
plane, which can be measured using single-molecule LOT pulling experiments,
shows that the thermodynamics of rupture as $f$ is increased, or the
assembly of the PK as $f$ is decreased, involves transitions between
the extended, intermediate, and the folded states. The predicted linear
relationship between $f_{\textrm{c}}$ and $\log C_{\textrm{m}}$
[Eq.\ref{eq:linear}] or $f_{\textrm{c}}\sim\sqrt{\log C_{\textrm{m}}}$
[Eq.\ref{eq:fc_sqrt}] shows that $\Delta\Gamma$ can be directly measured
in LOT experiments. The theoretical arguments leading to Eqs.(\ref{eq:linear})
and (\ref{eq:fc_sqrt}) are general. Thus, $f$ can be used as a perturbant
to probe ion-RNA interactions through direct measurement of $\Delta\Gamma$.

\section*{ACKNOWLEDGMENTS}

This work was completed when the authors were at the University of
Maryland. We are grateful to Jon Dinman, Xin Li, Pavel Zhuravlev,
Huong Vu, and Mauro Mugnai for comments on the manuscript. This work
was supported in part by a grant from the National Science Foundation
(CHE-1636424).

\bibliography{BWYVPK}


\clearpage

\setcounter{table}{0}
\setcounter{figure}{0}
\renewcommand{\thetable}{S\arabic{table}}
\renewcommand{\thefigure}{S\arabic{figure}}

\onecolumngrid

\begin{figure}
\includegraphics{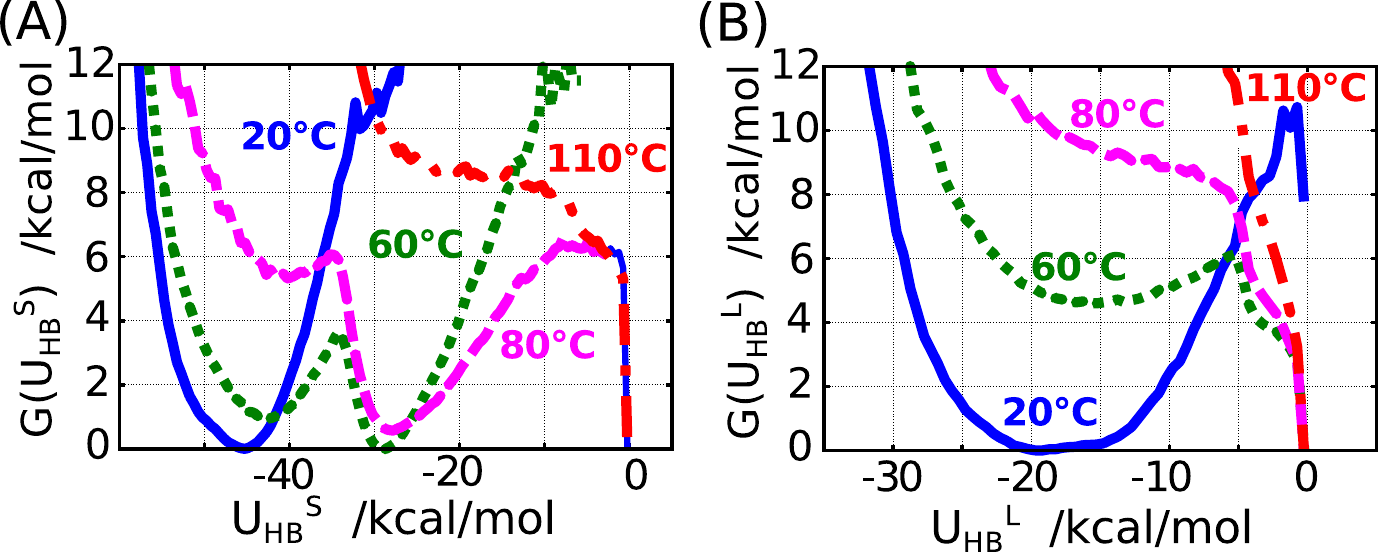}
\caption{\label{fig:UHB_500mM}Free energy profiles of the hydrogen bond energy of two
stems (A; $U_{\textrm{HB}}^{\textrm{S}}$) and two loops (stem-loop
interactions) (B; $U_{\textrm{HB}}^{\textrm{L}}$) at $C=500\textrm{ mM}$
and $f=0$. }
\end{figure}

\begin{figure}
\includegraphics{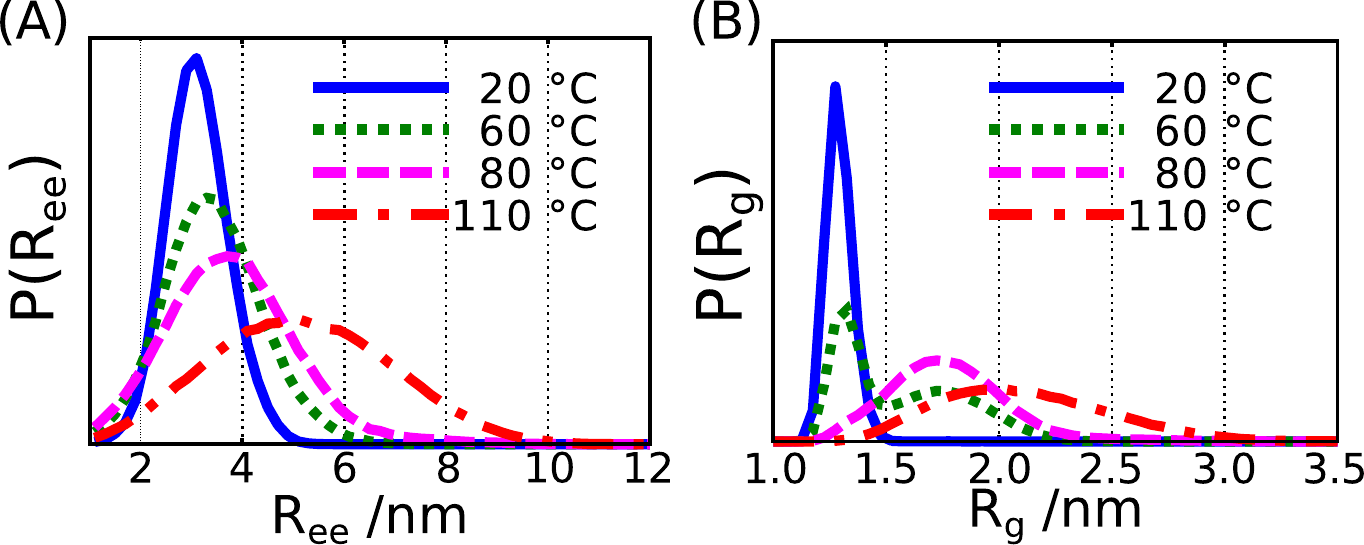}
\caption{\label{fig:ProbRgDee_500mM}Probability distributions of the molecular extension (A;
$R_{\textrm{ee}}$) and radius of gyration (B; $R_{\textrm{g}}$)
at $C=500\textrm{ mM}$ and $f=0$.}
\end{figure}

\begin{figure}
\includegraphics[scale=1.5]{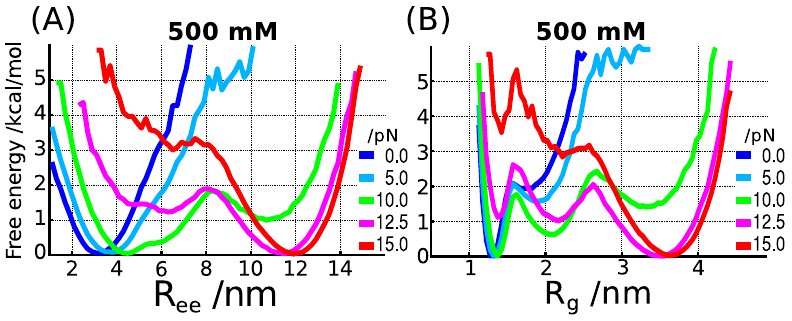}
\caption{\label{fig:GRgDee_500mM}Free energy profiles as a function of $R_{\textrm{ee}}$
(A) and $R_{\textrm{g}}$ (B) at 500 mM of monovalent salt.}
\end{figure}

\begin{figure}
\includegraphics[scale=1.5]{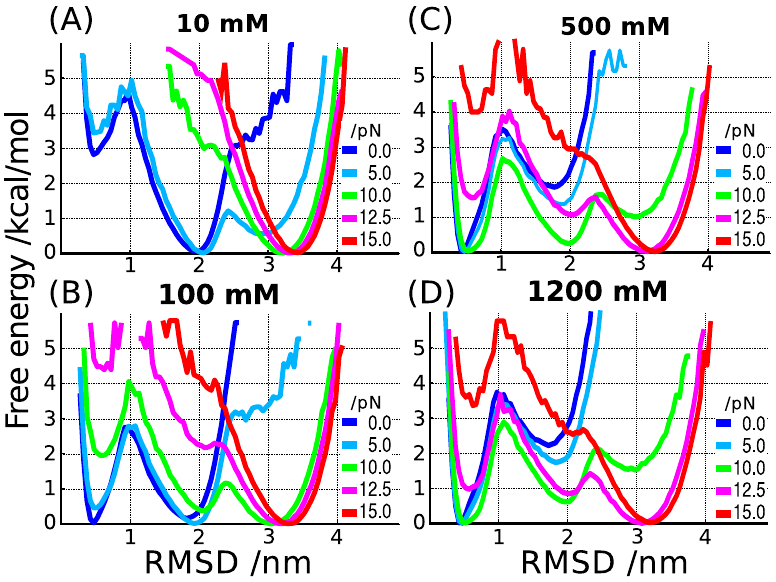}
\caption{\label{fig:RMSD}Free energy profiles as a function of RMSD at four different
salt concentrations, 10 mM (A), 100 mM (B), 500 mM (C), and 1200 mM
(D).}
\end{figure}

\begin{figure}
\includegraphics[scale=0.9]{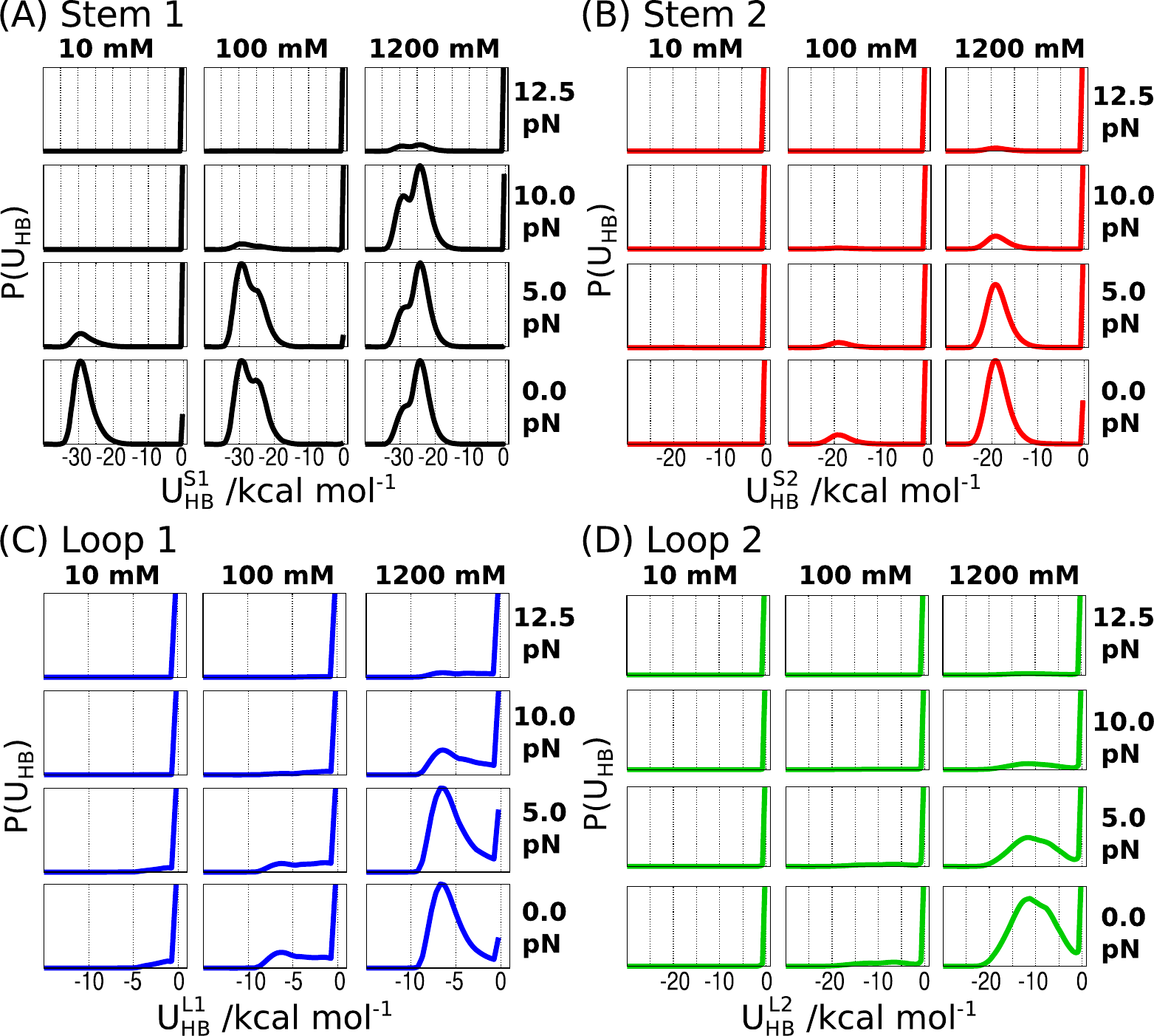}
\caption{\label{fig:Prob_HB_2nd}Probability distributions of hydrogen-bond energy. 
(A-D) Distributions of hydrogen bond energies for individual structural elements (A) Stem1;
(B) Stem 2; (C) Loop 1; and (D) Loop2. The salt concentrations and force
values are explicitly indicated.}
\end{figure}

\clearpage

\begin{table}
\caption{List of hydrogen bonding interactions in BWYV PK. First
8 pairs correspond canonical base pairs forming stems S1 and S2. Other
15 pairs are all tertially interactions forming either L1 or L2. Since
we use the three-interaction-site model, a letter ``s'', ``b''
or ``p'' is put following each nucleotide number to indicate one
of sugar, base, or phosphate, respectively.}
\begin{tabular}{ccc}
 & Nucleotide pair & Interaction type (atoms)\tabularnewline
\hline 
\hline 
1 & C3b - G18b & canonical b.p.\tabularnewline
2 & G4b -C17b & canonical b.p.\tabularnewline
3 & C5b - G16b & canonical b.p.\tabularnewline
4 & G6b - C15b & canonical b.p.\tabularnewline
5 & G7b - C14b & canonical b.p.\tabularnewline
6 & C10b - G28b & canonical b.p.\tabularnewline
7 & C11b - G27b & canonical b.p.\tabularnewline
8 & G12b - C26b & canonical b.p.\tabularnewline
\hline 
9 & G4b - A20b & N2 - N3\tabularnewline
10 & G4s - A20b & O2$^{\prime}$ - N1\tabularnewline
11 & C5b - A20s & O2 - O2$^{\prime}$\tabularnewline
12 & G6s - A21p & O2$^{\prime}$ - OP1\tabularnewline
13 & G7b - A24b & N2 - N1; N3 - N6\tabularnewline
14 & C8b - G12b & N4 - N7; N3 - O6\tabularnewline
15 & C8b - A25b & O2 - N6\tabularnewline
16 & C8b - C26b & O2 - N4\tabularnewline
17 & C14s - A25b & O2$^{\prime}$ - N1\tabularnewline
18 & C14b - A25b & O2 - N6\tabularnewline
19 & C15s - A23b & O2$^{\prime}$ - N1\tabularnewline
20 & C15b - A23b & O2 - N6\tabularnewline
21 & G16b - A21p & N2 - OP2\tabularnewline
22 & G16s - A21b & O2$^{\prime}$ - N7\tabularnewline
23 & C17N - A20s & O2 - O2$^{\prime}$\tabularnewline
\end{tabular}
\end{table}

\end{document}